\documentclass[a4paper,11pt]{article}
\pdfoutput=1
\usepackage{jheppub}

\usepackage{hyperref}
\hypersetup{
     colorlinks=true,       
     linkcolor=red,          
     citecolor=magenta,        
     filecolor=magenta,      
     urlcolor=blue           
}

\usepackage{graphicx}
\usepackage{placeins}
\usepackage{url}
\usepackage{multirow}
\usepackage{paralist}
\usepackage{amssymb}
\usepackage{amsmath}
\usepackage{xspace}
\usepackage{booktabs}
\usepackage[symbol]{footmisc}

\usepackage{lineno}

\newcommand{\newc}{\newcommand}
\newc{\cpp}{\textsf{C++}}
\newc{\HWS}{\textsf{Herwig 7}}
\newc{\HW}{\textsf{Herwig}}
\newc{\HSeven}{\textsf{H7}}
\newc{\HWHML}{\textsf{H7+HADML}}
\newc{\ThePEG}{\textsf{ThePEG}}
\newc{\HADML}{\textsf{HADML}}
\newc{\HWPPClass}[1]{\mbox{\href{http://projects.hepforge.org/herwig/doxygen/classHerwig_1_1#1.html}{\textsf{#1}}}}
\newc{\NC}{N_{\mathrm{c}}}

\begin{document}
\newcommand{\xju}[1]{\textcolor{red}{#1}}

\title{Towards a Deep Learning Model for Hadronization}

\affiliation[a]{Department of Physics and Astronomy, University of California, Irvine, CA 92697, USA}
\affiliation[b]{Physics Division, Lawrence Berkeley National Laboratory, Berkeley, CA 94720, USA}
\affiliation[c]{Berkeley Institute for Data Science, University of California, Berkeley, CA 94720, USA}
\affiliation[d]{Jagellonian University, Krakow, Poland}
\author[a,b]{Aishik Ghosh,}
\author[b]{Xiangyang Ju,}
\author[b,c]{Benjamin Nachman,}
\author[d]{and Andrzej Siodmok}

\abstract{
Hadronization is a complex quantum process whereby quarks and gluons become hadrons.  The widely-used models of hadronization in event generators are based on physically-inspired phenomenological models with many free parameters.  We propose an alternative approach whereby neural networks are used instead.  Deep generative models are highly flexible, differentiable, and compatible with Graphical Processing Unit (GPUs). We make the first step towards a data-driven machine learning-based hadronization model by replacing a component
of the hadronization model within the \HW\ event generator (cluster model) with a Generative Adversarial Network (GAN).  We show that a GAN is capable of reproducing the kinematic properties of cluster decays.  Furthermore, we integrate this model into \HW\ to generate entire events that can be compared with the output of the public \HW\ simulator as well as with $e^+e^-$ data.
}

\maketitle

\section{Introduction}
\label{sec:intro}

Simulations are essential tools for nearly all aspects of data analysis at particle colliders (see e.g., Ref.~\cite{Buckley:2011ms}).  These simulations are rooted in particle and nuclear physics and must model a large range in energy scales.  At the smallest distance scales, various forms of perturbation theory offer accurate, first-principles descriptions of hard-scatter particle reactions and collinear parton shower radiation.  The conversion from quarks and gluons to hadrons is performed using hadronization models.  Such approaches are physically inspired but are ultimately phenomenological models with many parameters that must be fit to data.  There are currently two main hadronization models, each inspired by a different description of strong dynamics in the low-energy region.  The linear confining potential motivated the string model~\cite{Andersson:1983ia,Sjostrand:1984ic} implemented in \textsf{Pythia}~\cite{Sjostrand:2007gs,Sjostrand:2006za} and preconfinement~\cite{Amati:1979fg,Bassetto:1979vy} inspired the cluster model~\cite{Webber:1983if} in \HW~\cite{Corcella:2000bw,Bahr:2008pv,Bellm:2015jjp,Bellm:2019zci} and \textsf{Sherpa}~\cite{Gleisberg:2008ta,Sherpa:2019gpd}.   In both models, there is an intermediate object between quarks/gluons and hadrons.
This intermediate object (string or cluster) takes as input the kinematic and flavor information from quarks and gluons and then has an approximately universal fragmentation into different hadron species that carry some fraction of the object's momentum.  

While existing hadronization models have been used successfully in a large number of phenomenological and experimental studies at the Large Hadron Collider and beyond, there is also significant room for innovation.  Existing models are not flexible enough to describe all of the properties of hadronization (see e.g. Ref.~\cite{2004.03540}).  Even so, these models still have a large number of parameters that need to be fit to data, which are adjusted (`tuned') using semi-automated programs like Professor~\cite{Buckley:2009bj}.  Existing tuning methods are not able to process high-dimensional observables or simultaneously tune many parameters because they rely on relatively simple \textit{surrogate models} to approximate the dependence of the data on the model.  A number of recently proposed automated tuning approaches employ sophisticated surrogate models~\cite{Ilten:2016csi,Andreassen:2019nnm,Wang:2021gdl}, but they all still require approximating complex relationships in high dimensions and therefore often are limited to relatively low-dimensional parameter spaces.

One natural alternative to the existing hadronization simulations is deep generative modeling.  Machine learning-based generators are highly flexible and differentiable by construction, which can aid parameter tuning. Three standard approaches to deep generative models include Generative Adversarial Networks (GANs)~\cite{Goodfellow:2014:GAN:2969033.2969125,Creswell2018}, (Variational) Autoencoders (VAEs)~\cite{kingma2014autoencoding,Kingma2019}, and Normalizing Flows (NFs)~\cite{10.5555/3045118.3045281,Kobyzev2020}.  While first proposed in high energy physics (HEP) to emulate an entire parton shower~\cite{deOliveira:2017pjk} or detector simulations~\cite{Paganini:2017dwg,Paganini:2017hrr}, deep generative models have now been proposed for many aspects of HEP simulations including matrix element generation~\cite{Bendavid:2017zhk,Butter:2019cae,Alanazi:2020jod,Bothmann:2020ywa,Gao:2020zvv,Gao:2020vdv,Butter:2021csz}, parton showers~\cite{deOliveira:2017pjk,Monk:2018zsb,Carrazza:2019cnt,Kansal:2020svm,Lai:2020byl,Kansal:2021cqp,Orzari:2021suh,Tsan:2021brw,Touranakou:2022qrp}, detector simulation~\cite{Paganini:2017hrr,Paganini:2017dwg,Vallecorsa:2019ked,SHiP:2019gcl,Chekalina:2018hxi,ATL-SOFT-PUB-2018-001,Carminati:2018khv,Vallecorsa:2018zco,Musella:2018rdi,Erdmann:2018kuh,Deja:2019vcv,Derkach:2019qfk,Erdmann:2018jxd,Oliveira:DLPS2017,deOliveira:2017rwa,Hooberman:DLPS2017,Belayneh:2019vyx,buhmann2020getting,2009.03796,Maevskiy:2020ank,deja2020endtoend,1816035,Rehm:2021zow,Rehm:2021zoz,Rehm:2021qwm,Khattak:2021ndw,Anderlini:2021qpm,Fanelli:2019qaq,Lu:2020npg,Buhmann:2021caf,Krause:2021ilc,Hariri:2021clz,Buhmann:2021lxj,ATLAS:2021pzo,Krause:2021wez,Bieringer:2022cbs}, and more (see Ref.~\cite{Butter:2022rso,Butter:2020tvl,Feickert:2021ajf} for reviews).  Using neural networks for modeling non-pertrubative inputs has a long history in the context of Parton Distribution Functions (PDFs) (Ref.~\cite{Forte:2002fg} through Ref.~\cite{Ball:2021leu}).  Similarly to hadronization models, PDFs cannot be calculated using perturbation theory.  In contrast to hadronization, PDFs are modeled as deterministic functions that are evolved in energy scale using perturbation theory~\cite{Gribov:1972ri,Dokshitzer:1977sg,Altarelli:1977zs}.

On the path towards a fully flexible, data-optimized, machine learning-based hadronization model, we demonstrate the first step by training a GAN to mimic a component of the cluster hadronization implementation in \HW. In particular, we replace part of the cluster decayer inside \HW\ with a GAN using the Open Neural Network Exchange (ONNX)~\cite{bai2019} interface to call the neural network inside the \texttt{C++} code.  This GAN-based cluster decayer, \HADML, is trained on \HW.  Future work will add additional complexity (cluster to cluster decays, color reconnection of clusters~\cite{Gieseke:2012ft,Gieseke:2019wcl,Bellm:2019wrh}, etc.) and will ultimately lead to a model that can be trained (tuned) on data.  This ultimate model will benefit from new, high-dimensional future measurements~\cite{Arratia:2021otl} that will provide the necessary constraining power for the flexible neural network approaches.

This paper is organized as follows.  Section~\ref{sec:methods} briefly introduces details of the \HW\ Monte Carlo event generator and how we interface a GAN in the hadronization stage.  Then, Sec.~\ref{sec:results} presents the first numerical results with the \HADML\ hadronization model.  The paper ends with conclusions and outlook in Sec.~\ref{sec:conclusion}.

\section{Methods}
\label{sec:methods}

\subsection{Dataset}
\label{sec:dataset}
The training data was created using the hadronization cluster model~\cite{Webber:1983if} .
The cluster model is based on t'Hooft's planar diagram theory \cite{tHooft:1973alw}:
the dominant color structure of Quantum Chromodynamics (QCD) diagrams in the perturbation expansion in $1/\NC$ can be represented in a planar form using color lines, which is commonly known as the limit $\NC \to \infty$. The
resulting color topology in Monte Carlo events with partons in the final-state
color features open color lines after the parton
showers. Following a non-perturbative isotropic decay of any left gluons in the parton jets to
quark-antiquark pairs, the event finally consists of color-connected partons in
color triplet or anti-triplet states. These parton pairs form color-singlet
clusters. This is so-called color preconfinement~\cite{Amati:1979fg}:
the tendency of the partons generated in the parton shower to be arranged in
color singlet clusters (pre-hadrons) with limited extension in both coordinate and momentum space.
  The principle of color preconfinement states that the mass distribution
of these clusters is independent of the hard-scattering process and its center-of-mass energy.
The cluster mass spectrum is not only universal but also peaked at low masses; therefore, most of the clusters decay into two hadrons and some just into one hadron.  However, there is a small fraction of clusters that are
too heavy for this to be a reasonable approach. Therefore, these heavy clusters are first split into
lighter clusters before they decay. Such decays of massive clusters are beyond the scope of this
publication, and we will consider it in future work.
Since the kinematics of a cluster decaying into a single hadron is trivial, our training data set only includes cases of decay into two hadrons. To further simplify the training data, we consider only decays into pairs of $\pi^{0}$.
Each decay in our data set was described with the following information:
the four-momentum of the cluster, the four-momenta of the two hadrons together with their flavor
(encoded as a Particle Data Group (PDG)~\cite{Tanabashi:2018oca} code), and the Pert flag.
Pert = 1 means that hadrons that contain a parton produced in the perturbative stage of the event remember the direction of the parton in the rest frame of the cluster.
To create the training data, we used $e^+e^-$ collisions at $\sqrt{s}=91.2$ GeV generated by
\HW\ version 7.2.1.
The only modification to the default generator settings was the
change that the hadrons produced from cluster decays were
on the mass shell\footnote{This setting can be achieved by adding the
command: \textsf{set ClusterDecayer:OnShell Yes} in the input file.}.

\subsection{GAN Model and Training}
We trained a conditional GAN to simulate the cluster decays. In a GAN, there is a Generator neural network (Generator for short) and a Discriminant neural network (Discriminator for short). Inputs to the Generator are the cluster's four vectors ($E$, $p_x$, $p_y$, $p_z$), and $N$ features sampled from a Gaussian distribution. The $N$ numbers are called \textit{noise}. $N$ is a hyperparameter and set to be 10. Outputs of the Generator are the polar angle, $\phi$, and azimuthal angle, $\theta$, of the leading hadron's momentum in the spherical coordinate system in the cluster frame, in which the two hadrons are created back-to-back. With the two angular variables, $\theta$ and $\phi$, and the cluster's four vector, we reconstruct the four vectors of the two outgoing hadrons as a postprocessing step. Inputs to the Discriminator are just the two angular variables coming from either the Generator, labeled as background, or those from the \HW, labeled as signal. The output of the Discriminator is a score that is higher for events from the \HW\ and lower for events from the Generator. The Discriminator is trained to separate signal from background. However, the Generator is trained to yield signal-like Discriminator score.

The GAN is based on multilayer perceptrons (MLPs). Both the Generator and the Discriminator are composed of a two-layer perceptron. Each perceptron consists of a sequence of \textsc{Keras}~\cite{keras} modules: a fully connected (dense) network of a hidden size of 256, a batch normalization layer, and a \textsc{LeakyReLU} activation function~\cite{leakyReLU}.  These parameters were not extensively optimized.

To help train a GAN, we preprocessed the training data. The incoming cluster's four vector is scaled so that their values are between -1 and 1; so are the two angular variables ($\phi$ and $\theta$). In this way, all inputs and outputs are within the same scale. Finally, we use the $\tanh$ activation function as the last layer of the Generator. The Discriminator and the Generator are trained separately and alternately by two independent \textsc{Adam} optimizers~\cite{adam}, both with a learning rate of $10^{-4}$, for about 1000 epochs.

\begin{figure}[htb]
    \centering
    \includegraphics[width=0.6\linewidth]{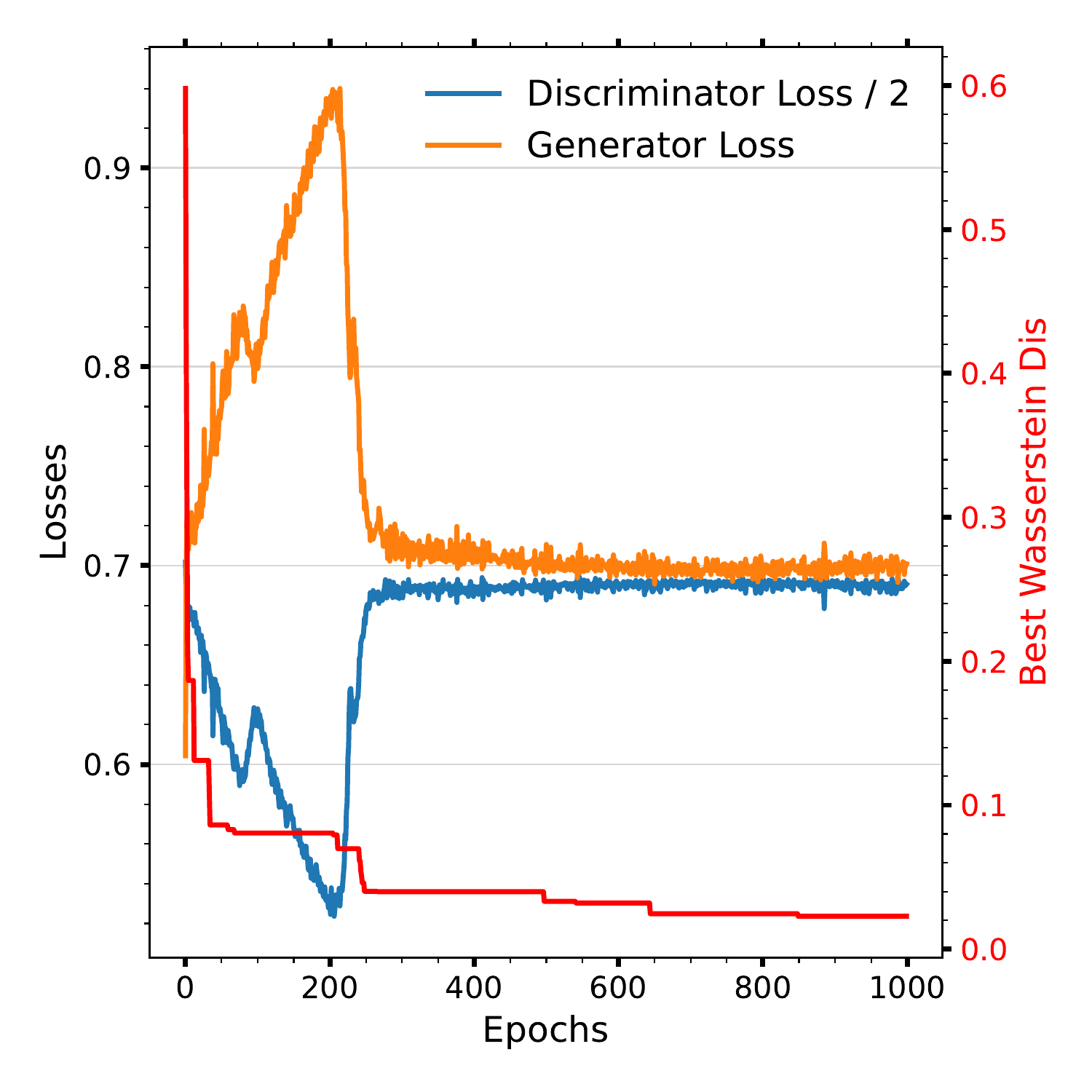}
    \caption{Generator loss and discriminator loss and progressive best Wasserstein distance as a function of the training epochs for training a GAN with events where two partons are with Pert = 0. Both Generator and Discriminator loss are the binary-crossentropy loss, and the Discriminator loss is divided by two for visualization purposes. The progressive Wasserstein distance is gauged by the right side of the $y$ axis.}
    \label{fig:train_loss_evolve}
\end{figure}
Figure~\ref{fig:train_loss_evolve} shows the evolution of the Discriminator loss, which is divided by two for visualization purposes, the Generator loss, and the progressive best total Wasserstein distances\footnote{This is a common metric in machine learning that quantifies the minimal `work' required to transform one density into another, where work, in this case, is defined as the integral of the density multiplied by the distance moved. }~\cite{Vaserstein,Kantorovich} for training a GAN with events where two partons are with Pert = 0. The total Wasserstein distance summing over the distances of all variables, is calculated after training for one epoch and only the smallest value is plotted. At the beginning of the training (epoch $< 70$), even though the Generator loss is going up, we see a rapid drop in the Wasserstein distance until the Generator loss is beyond 0.8. For more than 100 epochs, the Discriminator keeps outperforming the Generator as seen by the increasing Generator loss and the decreasing Discriminator loss. This situation is changed around epoch 200 and finally, the two networks reach an equilibrium around epoch 250. Beyond epoch 600, we only see about 0.002 improvements in the Wasserstein distance. The best model for events with partons of Pert = 0, is found at the epoch 849 with a total Wasserstein distance of 0.0228. A similar analysis was performed when training events with at least one parton with Pert = 1.
\subsection{Integration into Herwig}
Each part of \HW\ is implemented as a
  \cpp\ class that contains
  the implementation of the \HW\ physics models, inheriting from an
  abstract base class in \ThePEG~\cite{Lonnblad:2006pt}.
  The \HWPPClass{ClusterHandronizationHandler} is the class that controls the
  cluster hadronization model. Our ultimate goal will be to replace the entire ClusterHandronizationHandler with its ML counterpart. However, since in these studies, we concentrate on the decay of clusters into two hadrons, it was sufficient to modify \HWPPClass{ClusterDecayer} - a helper class of the ClusterHandronizationHandler
  that controls this process.
The generative model trained in Python using TensorFlow is converted into the ONNX format~\cite{bai2019} and integrated into the \HW\ chain using the \texttt{C++} API of ONNX Runtime~\cite{onnxruntime}. The advent of the ONNX format makes it possible to train a model in one software and hardware environment and then apply it in a completely different environment. ONNX Runtime is well suited for running fast neural network inference as part of a large \texttt{C++} workflow, and by using it, we avoid having to integrate and maintain \textsc{TensorFlow}~\cite{AbaBar16Tensorflow} within the \HW\ framework.

All preprocessing and postprocessing steps performed for training are repeated within \HW\ for inference. The entire simulation chain, including the GAN, is then run in \HW\ in order to produce the final comparisons and results.

\section{Results}
\begin{figure}
\centering
\includegraphics[width=0.49\textwidth]{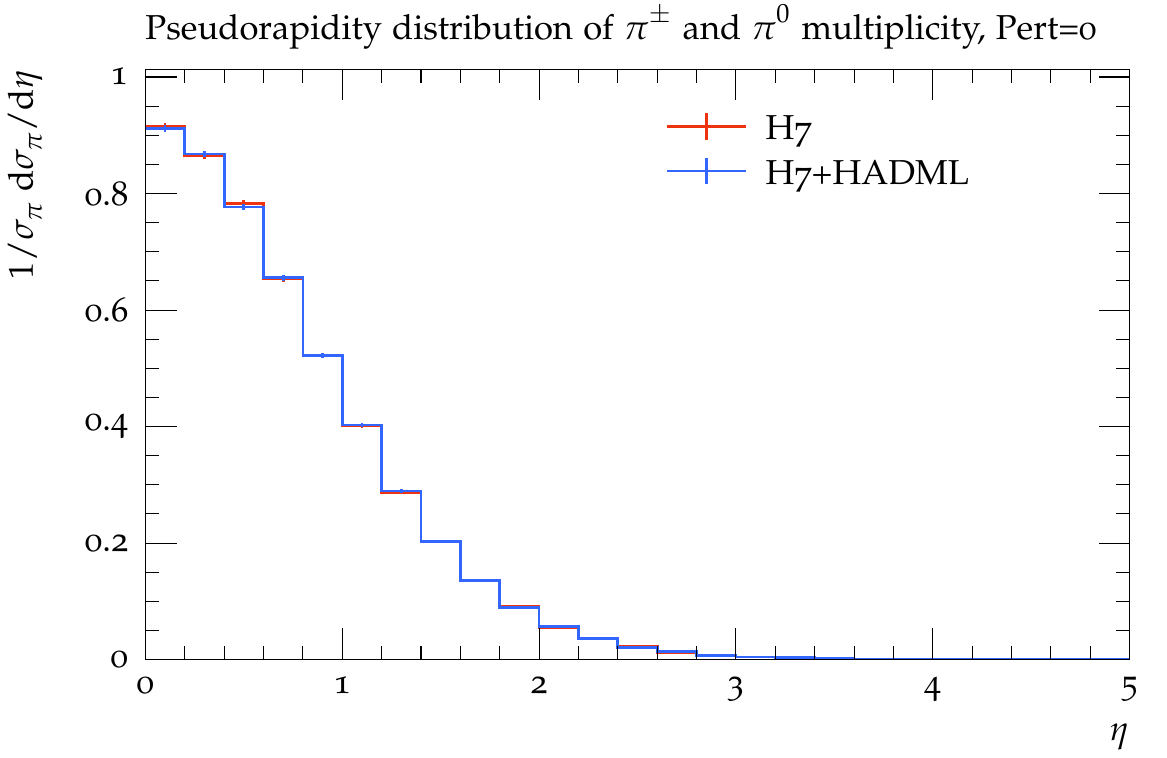}
\includegraphics[width=0.49\textwidth]{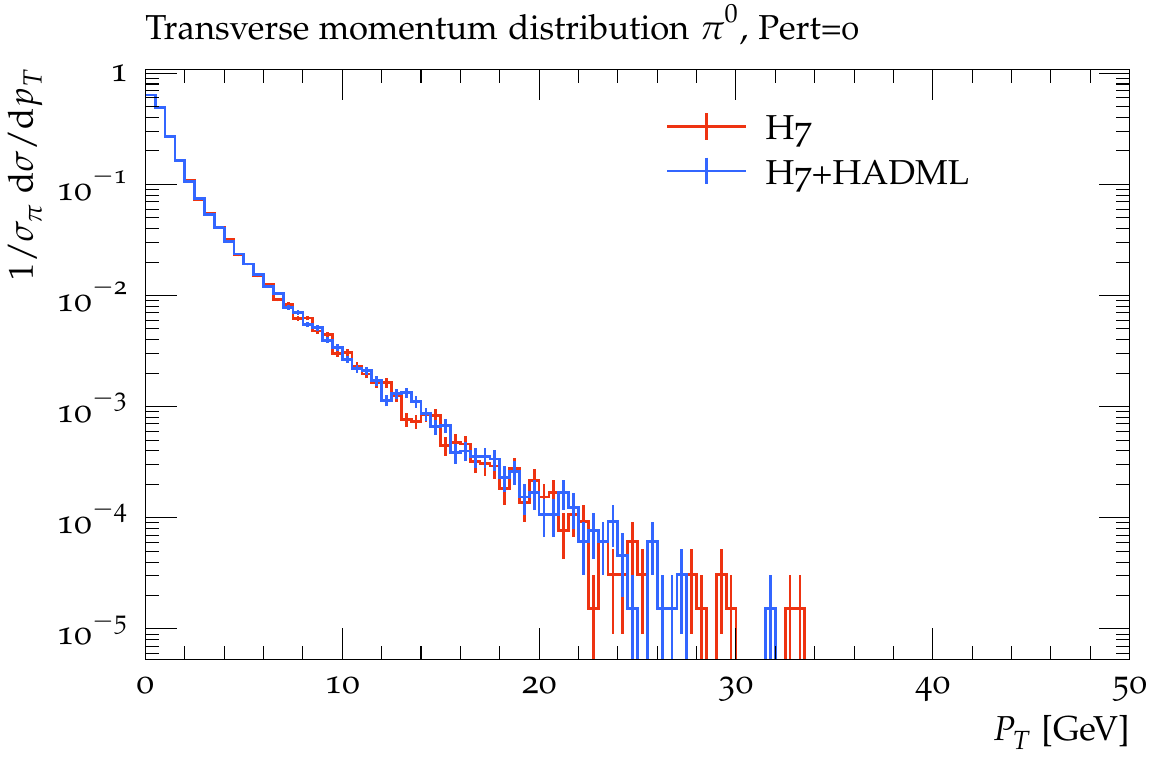}
\includegraphics[width=0.49\textwidth]{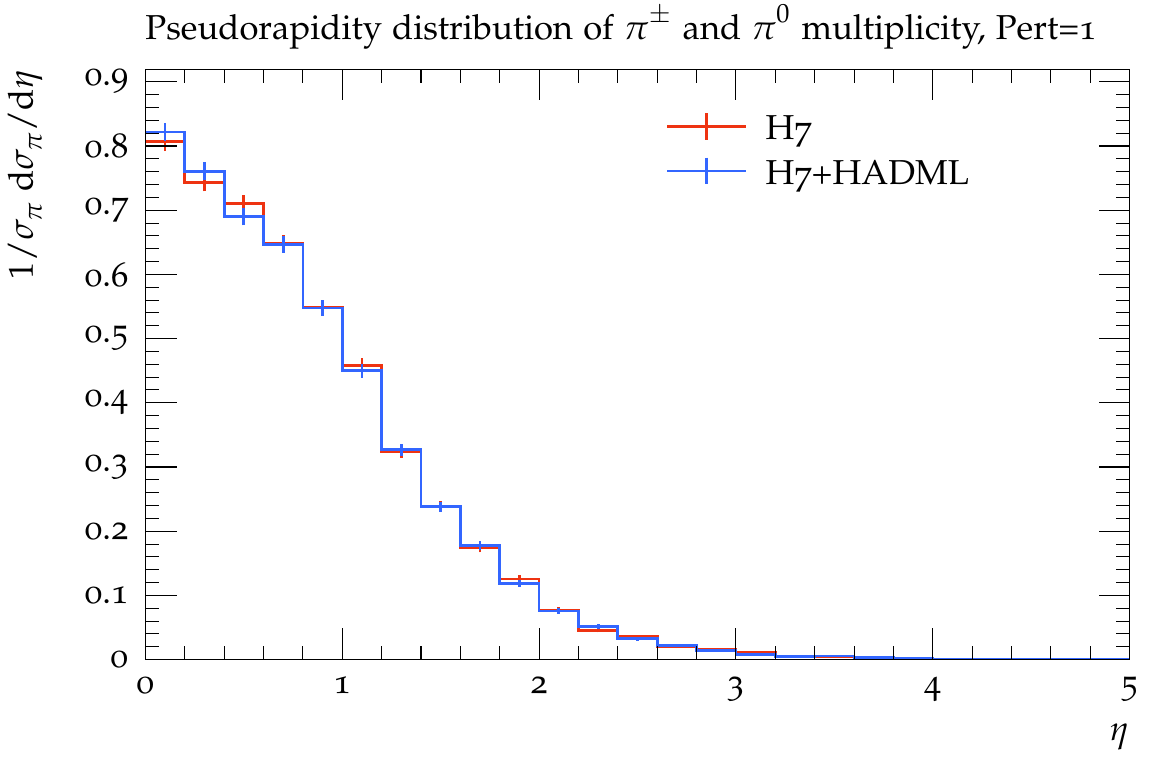}
\includegraphics[width=0.49\textwidth]{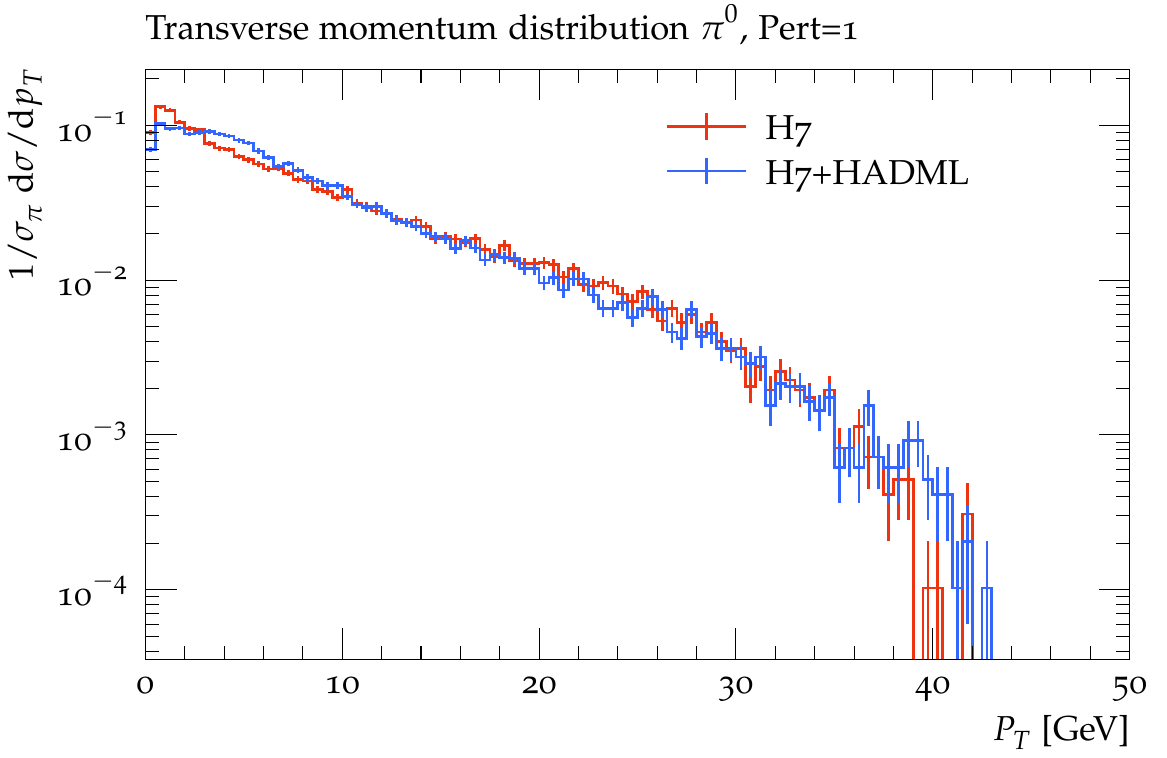}
    \caption{
    \label{fig:2Pi0}
  Pseudorapidity (left panels) and transverse momentum (right panels) distribution of $\pi^0$
  from decays of Pert=0 (upper panels) and Pert=1 (lower panels) clusters produced in $e^+e^-$
  collisions at $\sqrt{s}=91.2$~GeV.
    }
\end{figure}
\label{sec:results}
\begin{figure}[htb]
\centering
\includegraphics[width=0.49\textwidth]{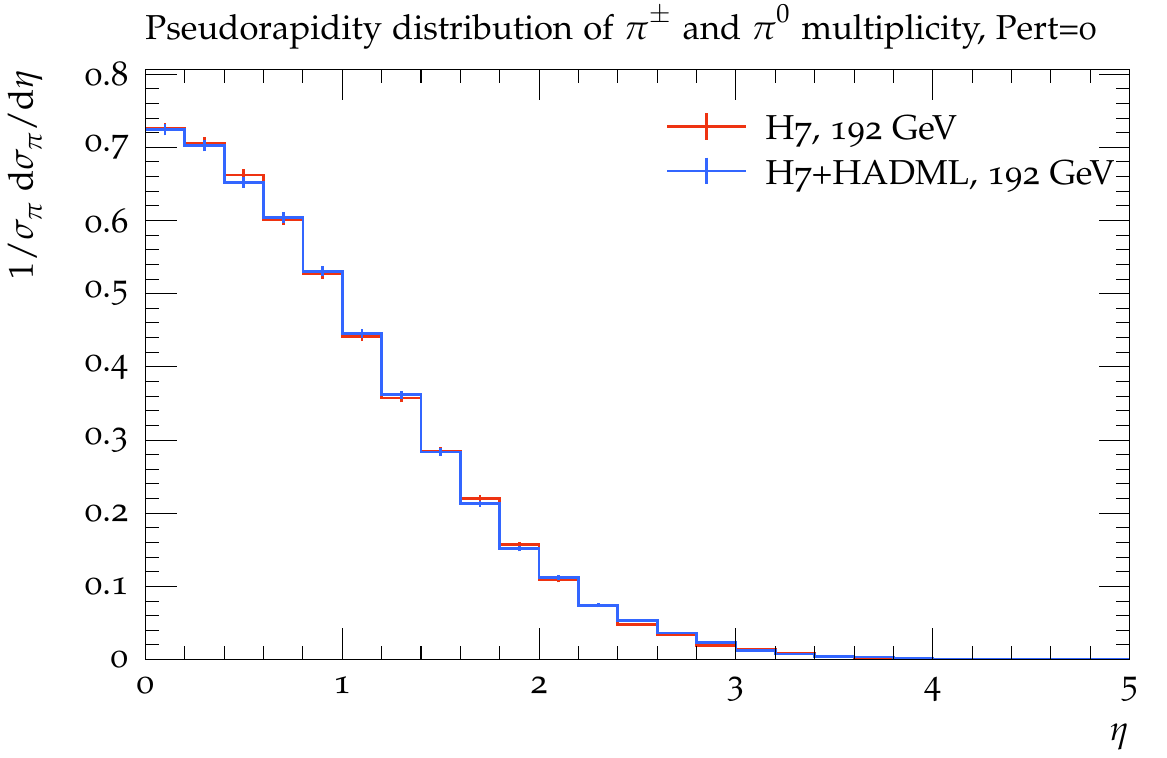}
\includegraphics[width=0.49\textwidth]{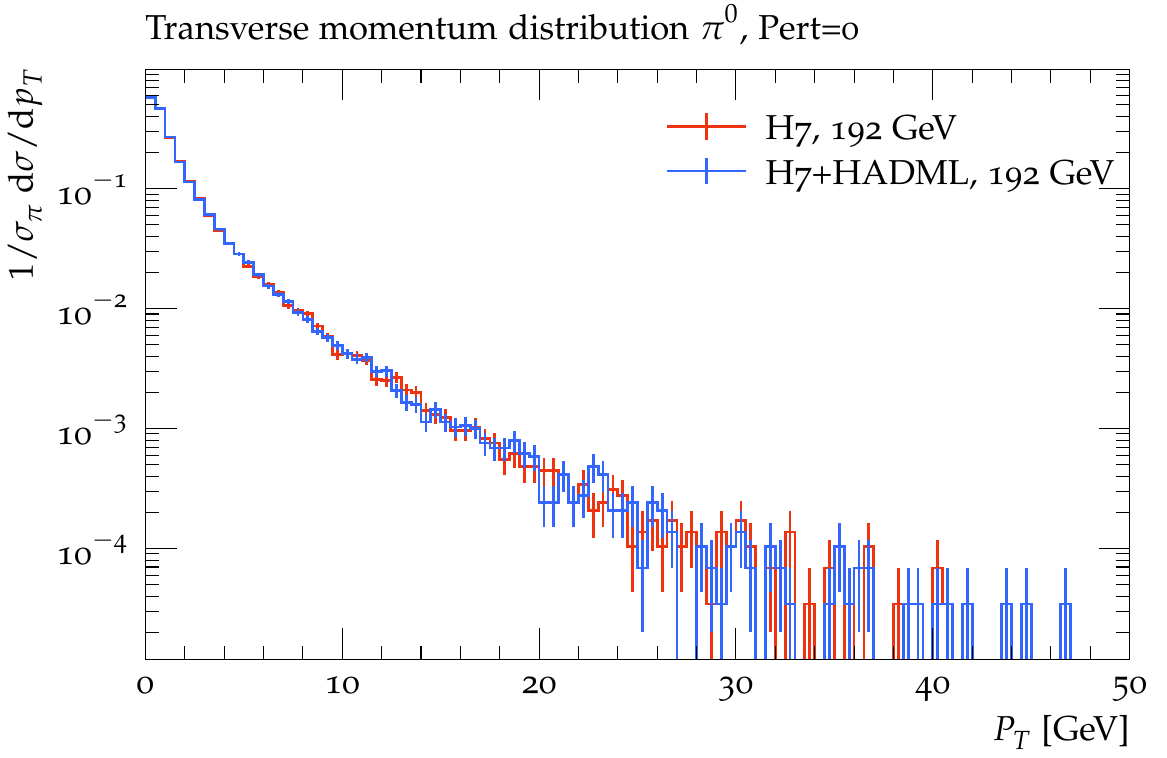}
\includegraphics[width=0.49\textwidth]{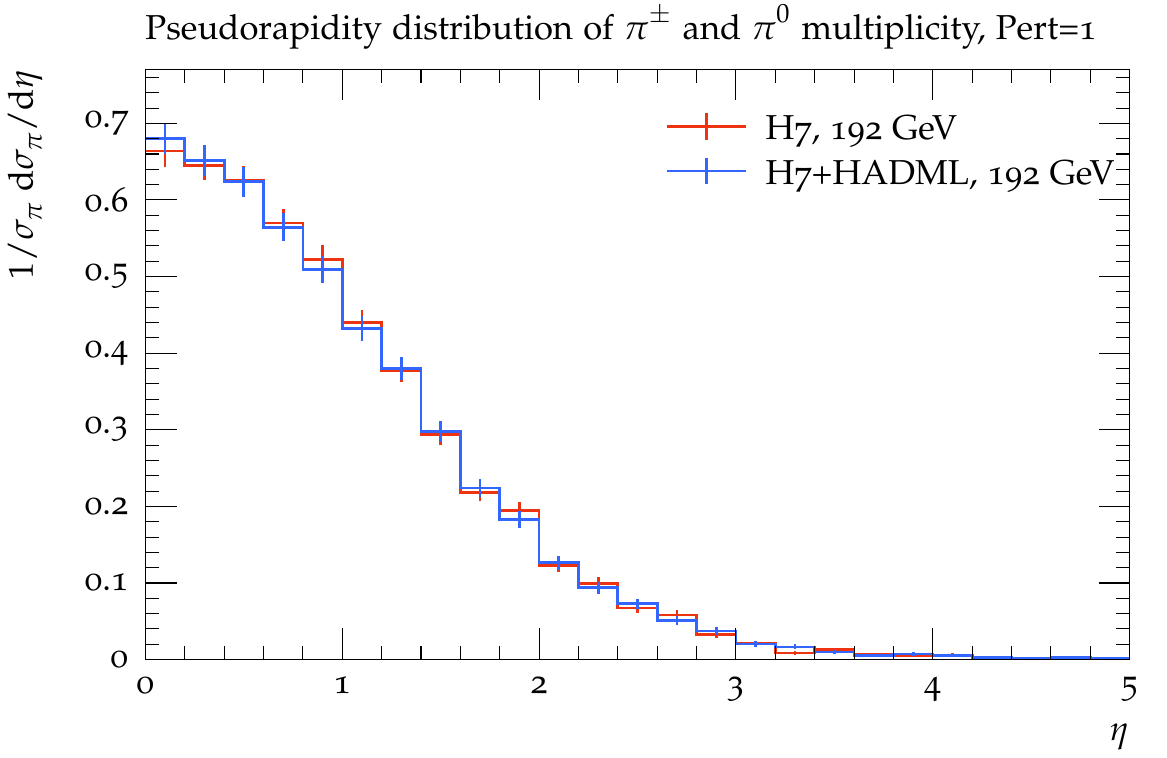}
\includegraphics[width=0.49\textwidth]{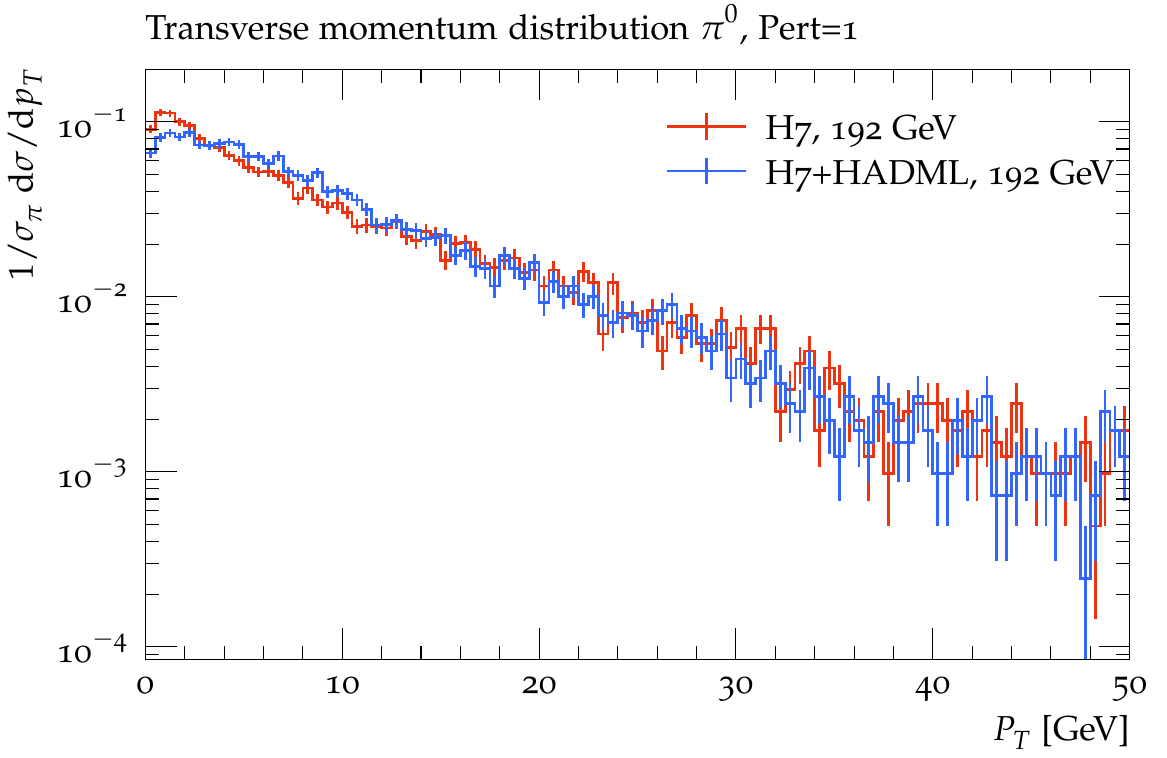}
    \caption{\label{fig:2Pi0192}  
 Pseudorapidity (left panels) and transverse momentum (right panels) distribution of $\pi^0$
  from decays of Pert=0 (upper panels) and Pert=1 (lower panels) clusters produced in $e^+e^-$ collisions at $\sqrt{s}=192$~GeV.  
    }
\end{figure}
Section~\ref{sec:lowlevel} provides low-level results of individual cluster decays while Sec.~\ref{sec:full} includes full event simulations and comparisons to $e^+e^-$ data.
\subsection{Low-level Validation}
\label{sec:lowlevel}

Since the training data contained only clusters produced
in $e^+e^-$ collisions at $\sqrt{s}=91.2$~GeV that decayed into $\pi^{0}$ pairs, we begin by comparing the $\pi^{0}$ kinematic variables generated by \HADML\ and \HW\ precisely in such decays.
The data generated by \HW, with which we compared the results of \HADML\ in this section,
were not used for training.
In Fig.~\ref{fig:2Pi0} we show the distribution of the pseudorapidity (left panels) and transverse momentum  distribution (right panels) of $\pi^0$ from the decays of the Pert = 0 (upper panels) and Pert=1 (lower panels) clusters.
As expected, we see that the transverse momentum spectra of pions coming from clusters containing ``perturbative'' quarks (Pert=1) are harder compared to those containing only non-perturbative partons
(Pert=0).
However, the most important observation from Fig.~\ref{fig:2Pi0} is that \HWS\ + \HADML\ (labeled on figures as \HWHML) matches the pseudorapidity of the pions generated by \HWS\ with the cluster model (labeled as \HSeven\ on figures). Transverse momentum spectra that extend over several orders of magnitude are also well approximated by \HWHML. Taking a closer look at these distributions, we see minor differences for low transverse momenta in the case of clusters that have a memory of perturbative quarks (bottom-left panel in Fig.~\ref{fig:2Pi0}). Such small differences are, of course, acceptable, especially since the information about the  four-momentum of partons that make up the clusters were not used for training. Taking this additional information into
account in the training process will likely eliminate these minor differences.
However, this is beyond the scope of this publication, and we will leave this problem for future work.

It is crucial that the hadronization model is universal, i.e., that it works independently of the hard process or collision energy. As we described in the Sec.~\ref{sec:dataset}  the cluster model has this property.
To test whether \HADML\ also is universal, we decided to repeat the comparison made at the beginning of this section, but this time generating events with collision energies twice as high as those used in the training data. In Fig.~\ref{fig:2Pi0192} we show
$\pi^{0}$ kinematic variables generated by \HWHML\ and \HWS\
in $e^+e^-$ collisions at $\sqrt{s}=192$~GeV.
We can see that all distributions are described very similarly by both models, which reassured us that the \HADML\ model is also universal.

\begin{figure}[htb]
\centering
\includegraphics[width=0.49\textwidth]{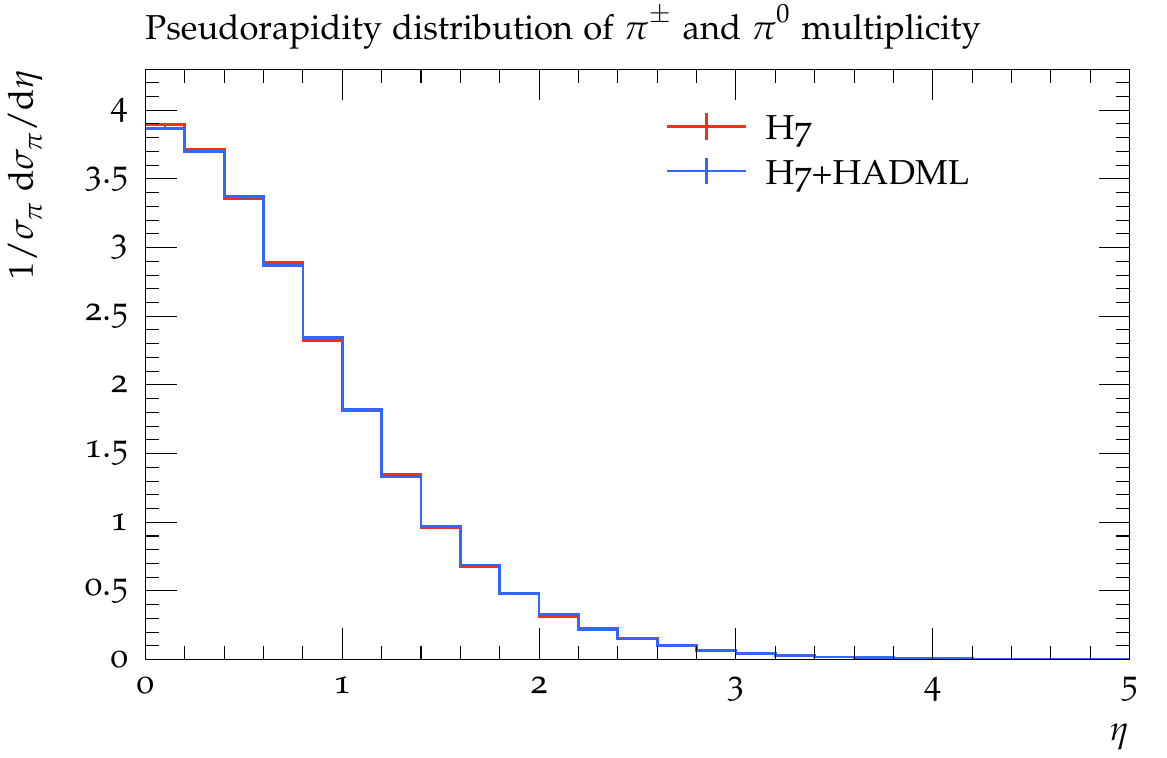}
\includegraphics[width=0.49\textwidth]{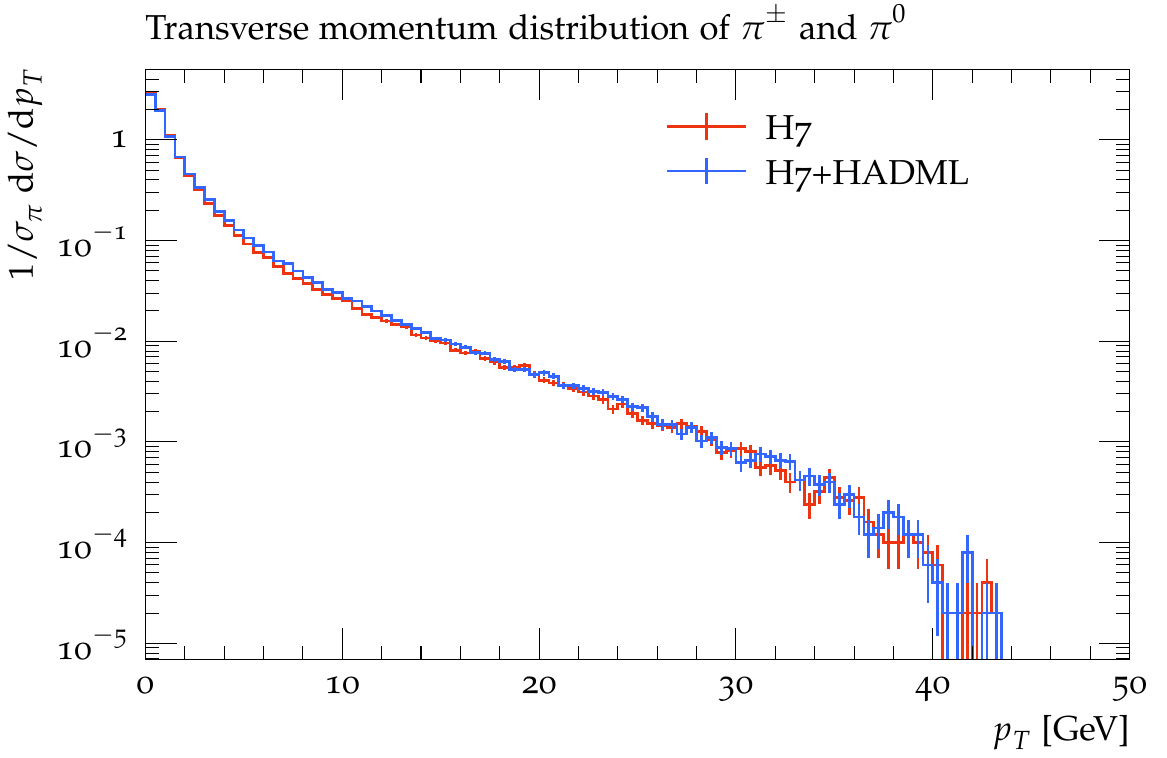}
\includegraphics[width=0.49\textwidth]{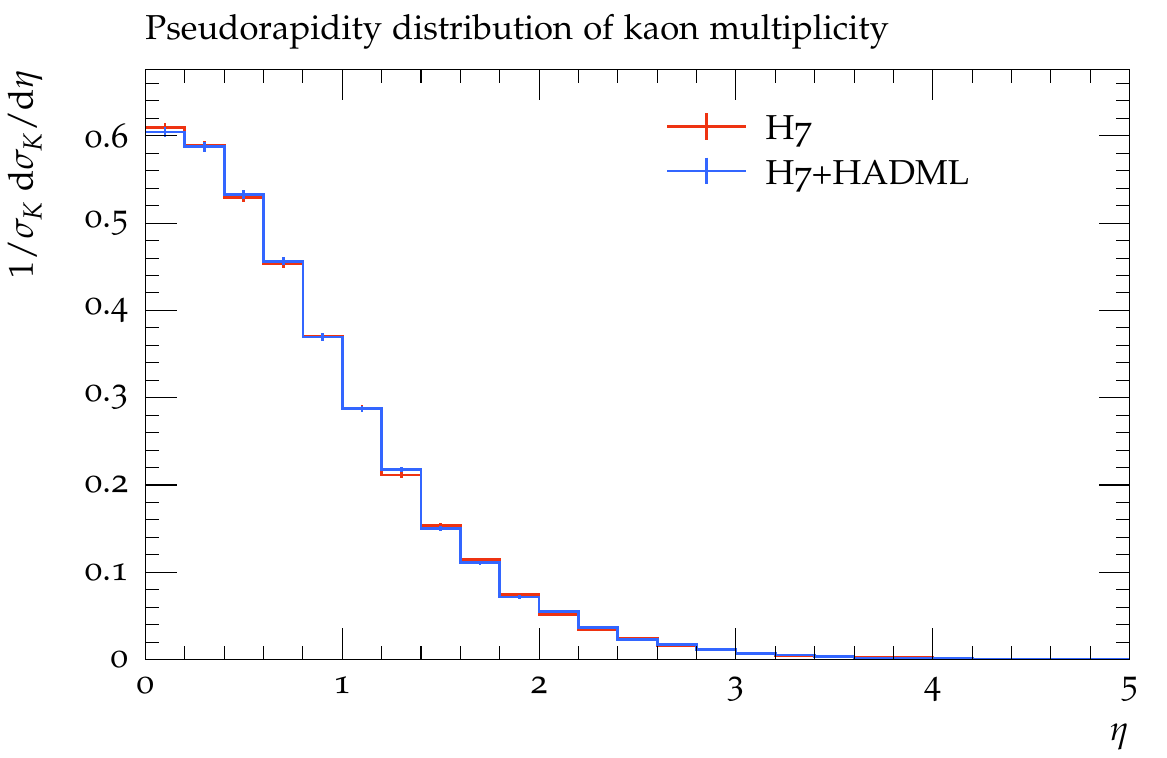}
\includegraphics[width=0.49\textwidth]{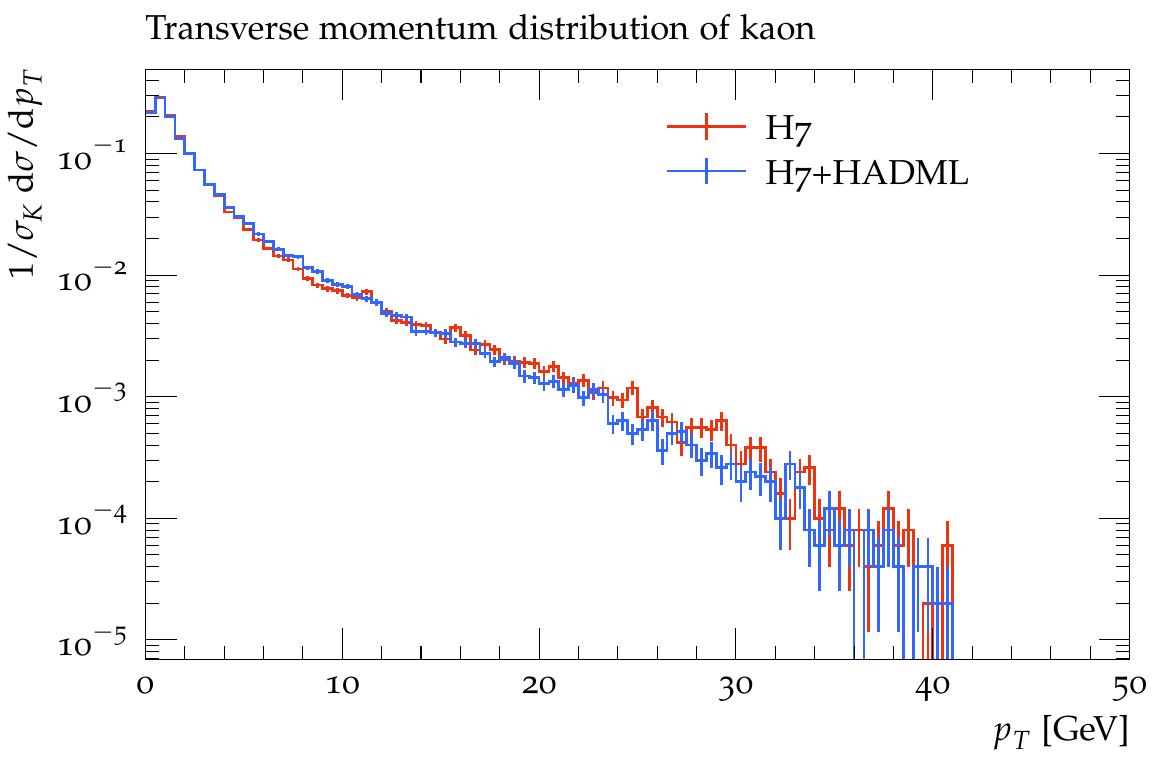}
\includegraphics[width=0.49\textwidth]{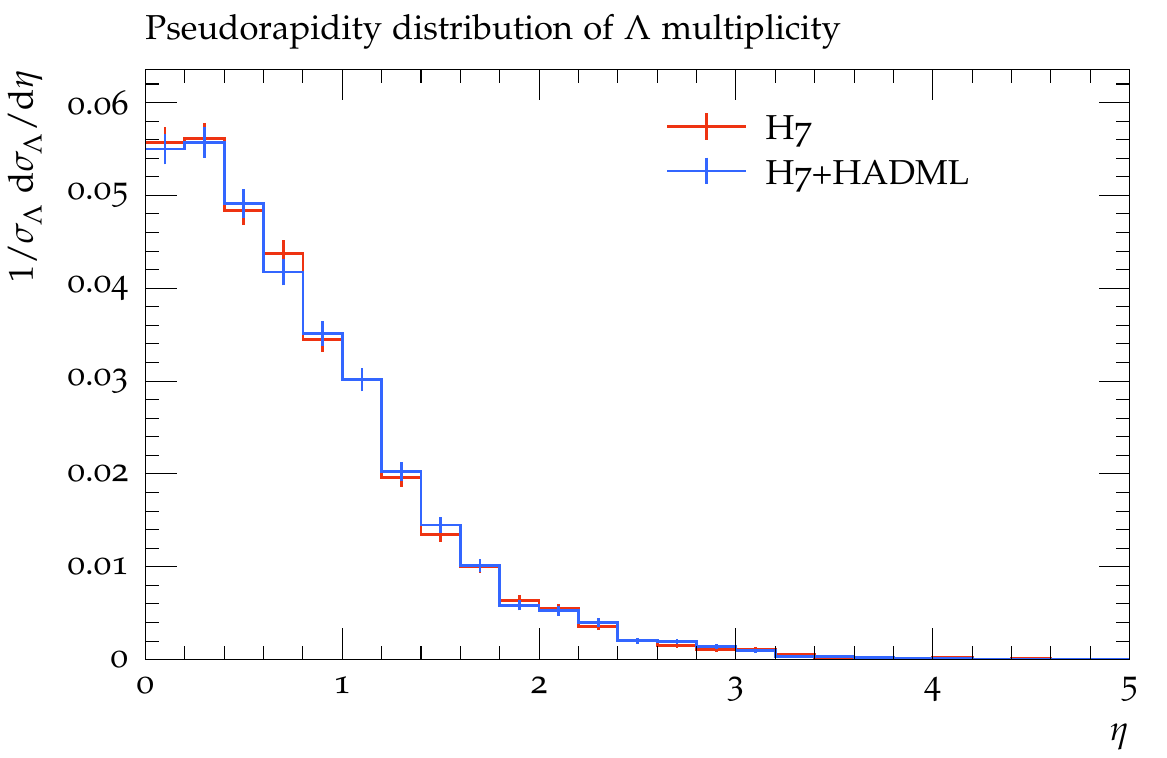}
\includegraphics[width=0.49\textwidth]{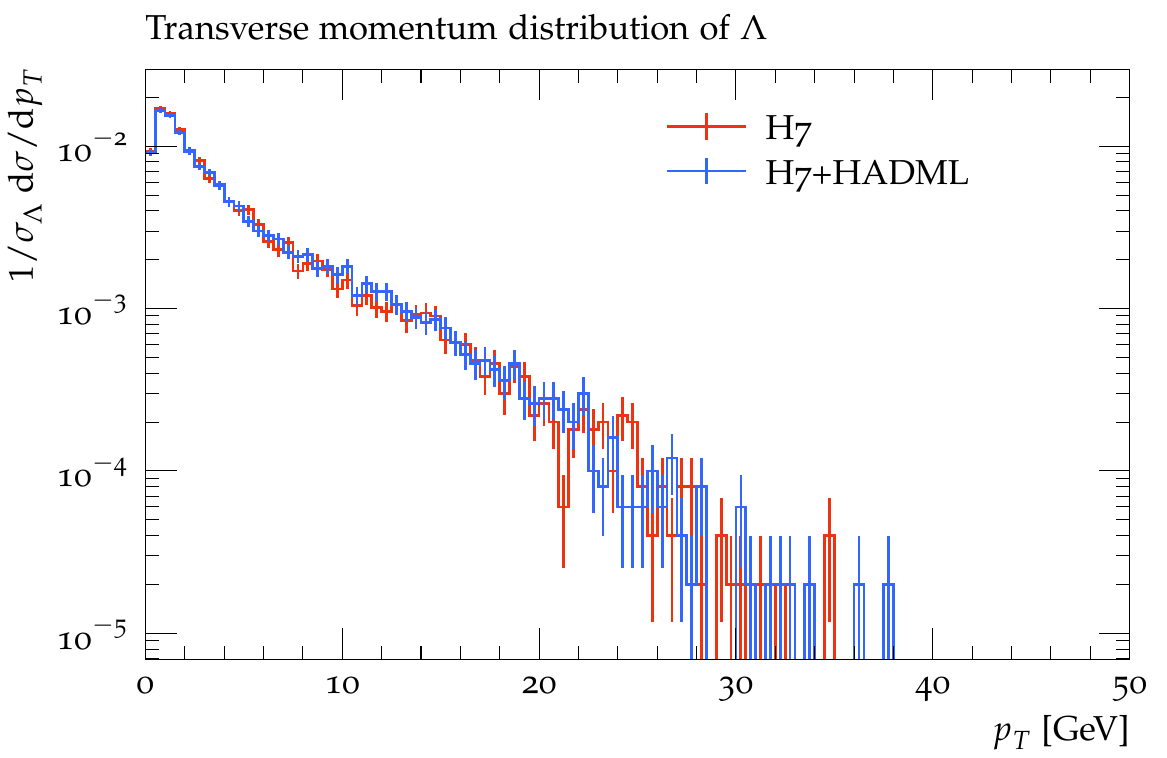}
    \caption{\label{fig:Hadrons}  
    Pseudorapidity (left panels) and transverse momentum (right panels) distribution of $\pi^{\pm}$ and $\pi^0$ (first row), Kaons (second row) and Lambdas (third row).  
}
\end{figure}

The last thing we need to check before using \HADML\ to simulate the decay of all clusters into hadron
pairs in \HW\ is whether the model is able to describe the kinematics of
other hadrons than $\pi^{0}$. In Fig.~\ref{fig:Hadrons} we present the pseudorapidity (left panels) and transverse momentum (right panels) distribution of $\pi^{\pm}$ and $\pi^0$ (first row), kaons (second row) and lambdas (third row). We see that the distributions differ for the various hadrons, but they are all described almost identically by both models. This encouraged us to perform a comparison with experimental data in which the kinematics of all hadrons\footnote{Except for a small number of hadrons that come from the decay of a cluster into a single hadron for which the kinematics is trivial.} in \HW\ are generated by \HADML\ model.   
\subsection{Full-event Validation}
\label{sec:full}

In this section, we generate full events using \HADML\ integrated into \HW\ and compare the results also to data from LEP\footnote{Note that the data are for illustration only - given that the GAN is trained on \HW, we cannot expect it to outperform \HW.  Tuning to data is a longer-term goal of this research (see Sec.~\ref{sec:conclusion}).}.  In particular, we consider an analysis from DELPHI with data collected at $\sqrt{s}=91.2$ GeV~\cite{DELPHI:1996sen} using RIVET\footnote{\url{https://rivet.hepforge.org/analyses/DELPHI_1996_S3430090}.}~\cite{Buckley:2010ar}.  These events correspond to hadronic $Z$ boson decays with a number of event shape and identified hadron spectra.  These data have been used for hadronization parameter tuning~\cite{DELPHI:1996sen,Reichelt:2017hts}.  

\begin{figure}[htb]
\centering
\includegraphics[width=0.49\textwidth]{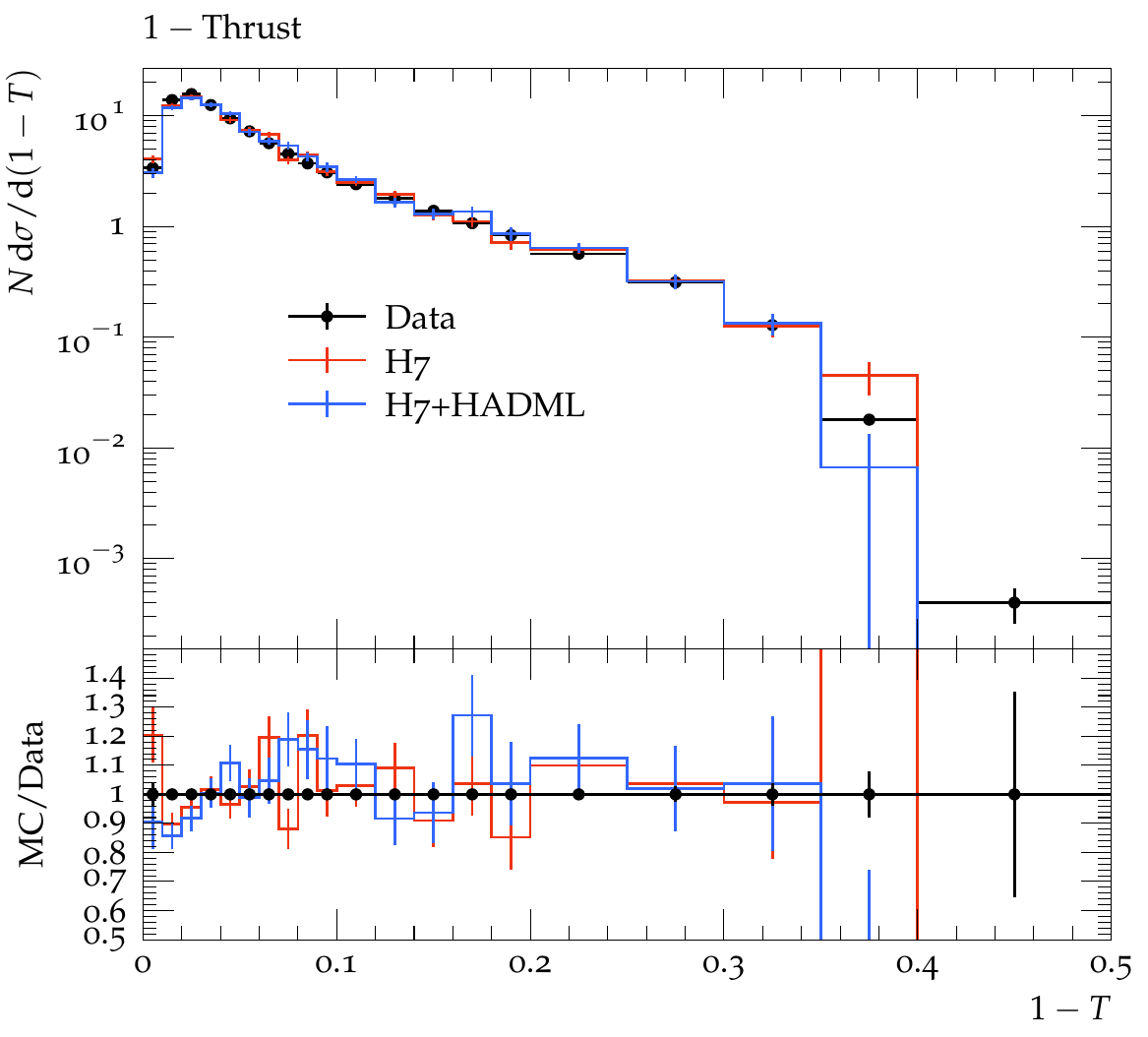}
\includegraphics[width=0.49\textwidth]{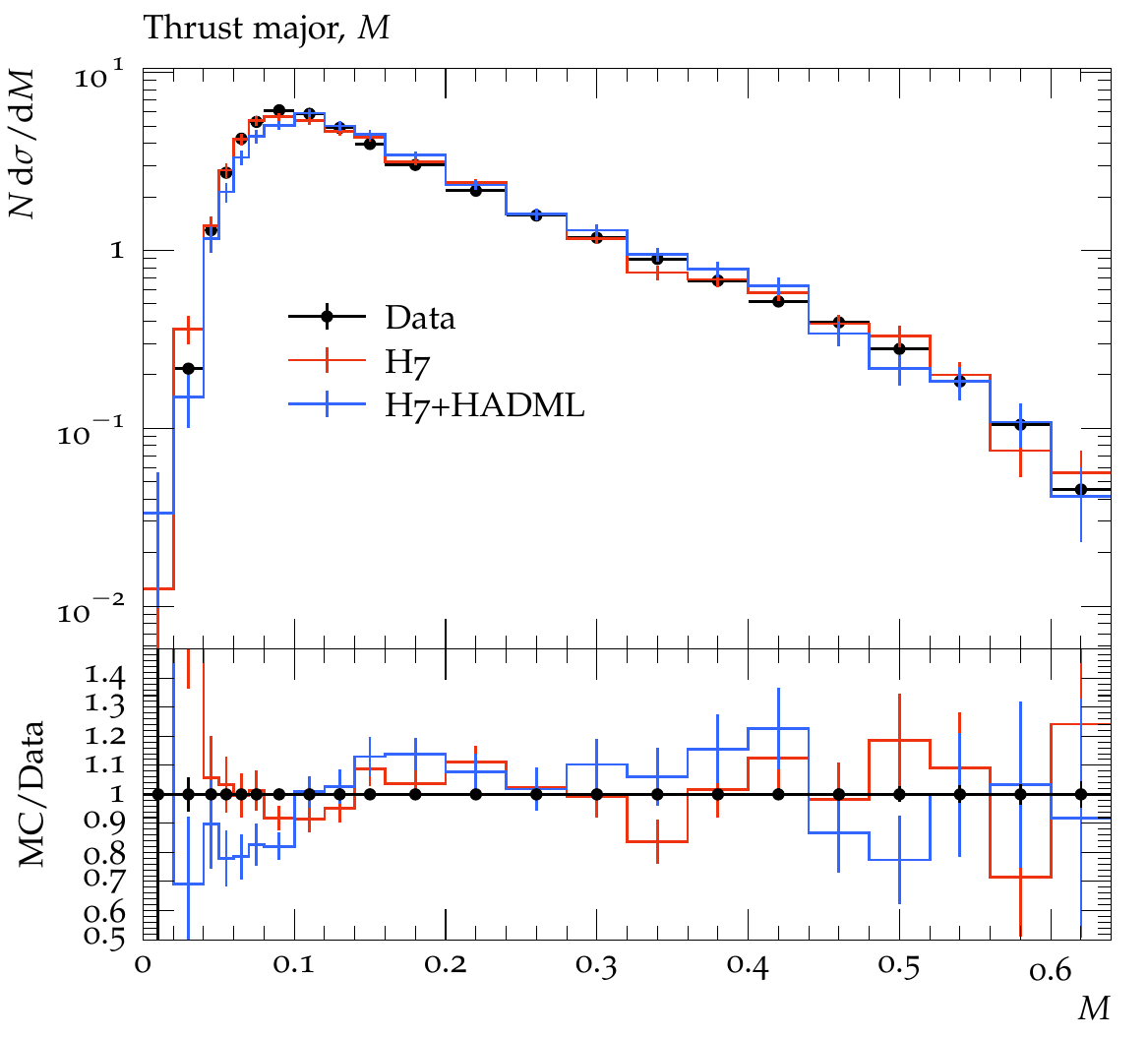}
\includegraphics[width=0.49\textwidth]{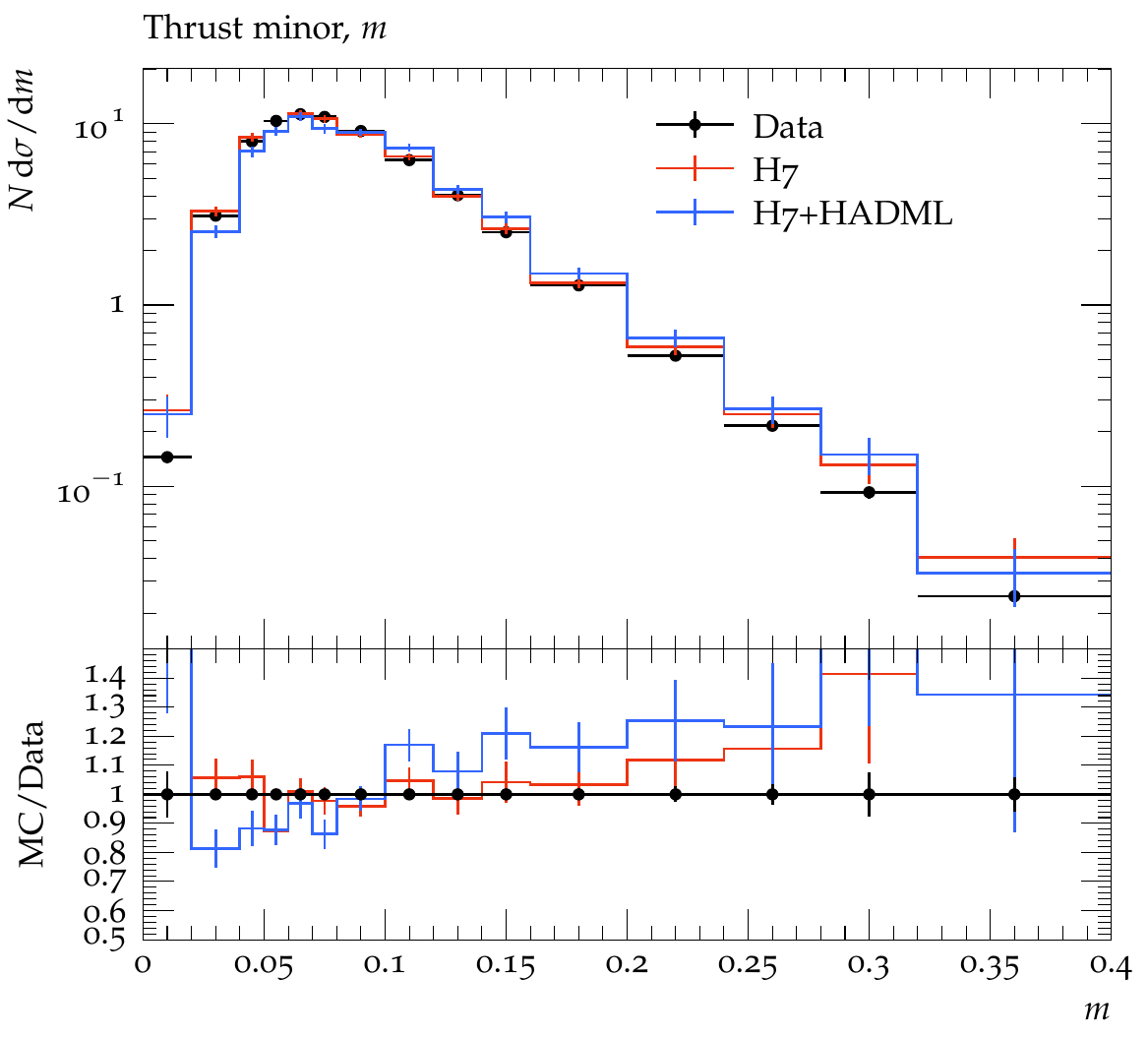}
\includegraphics[width=0.49\textwidth]{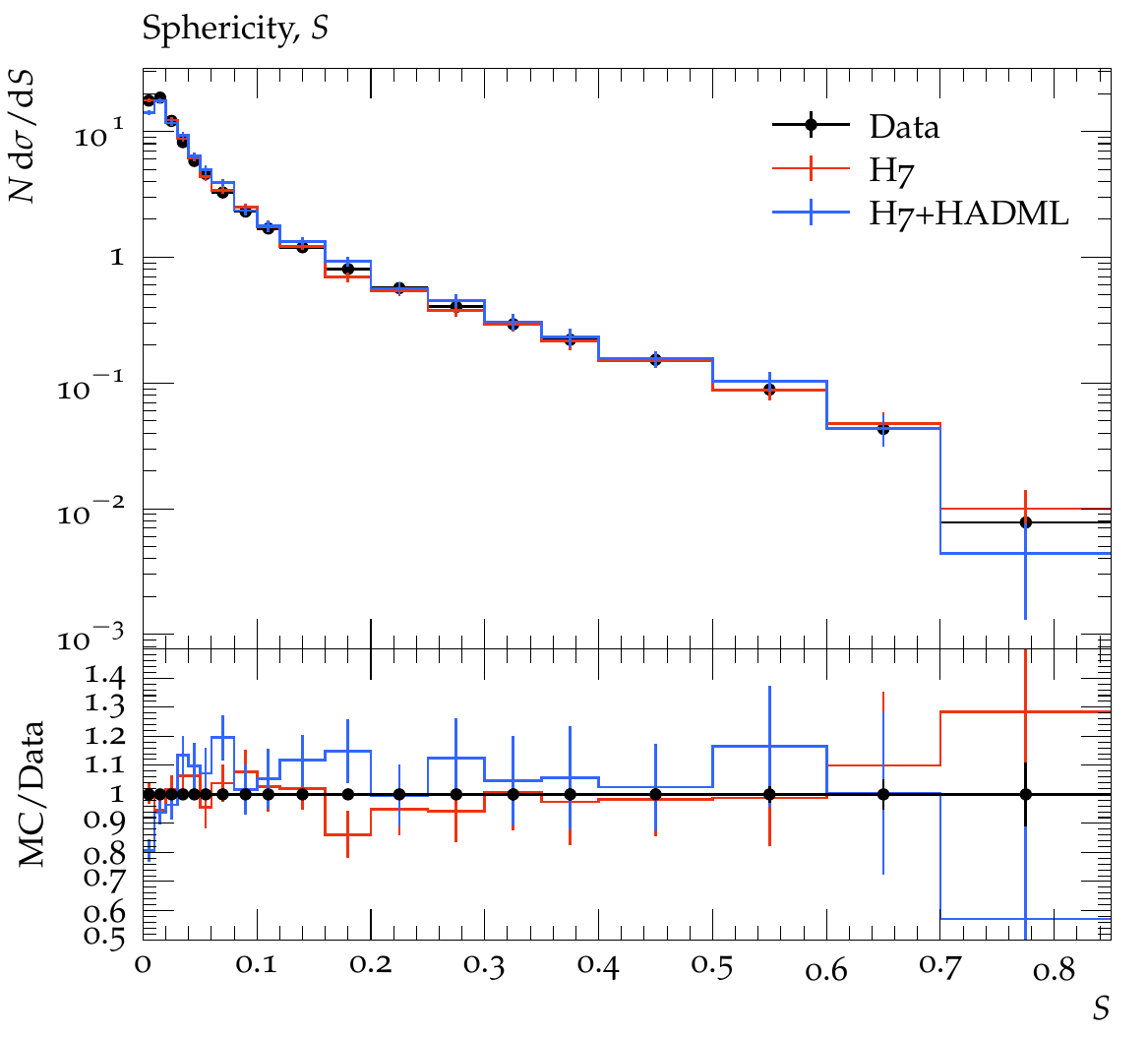}
    \caption{\label{fig:thrust}  
  Normalized, differential cross-sections of Thurst (top left), Thrust major (top right), Thrust minor (lower left), and Sphericity (lower right) for \HW, \HW\ with \HADML, and for data from DELPHI at LEP.  Error bars on the predictions represent statistical uncertainties.
    }
\end{figure}

\begin{figure}[htb]
\centering
\includegraphics[width=0.49\textwidth]{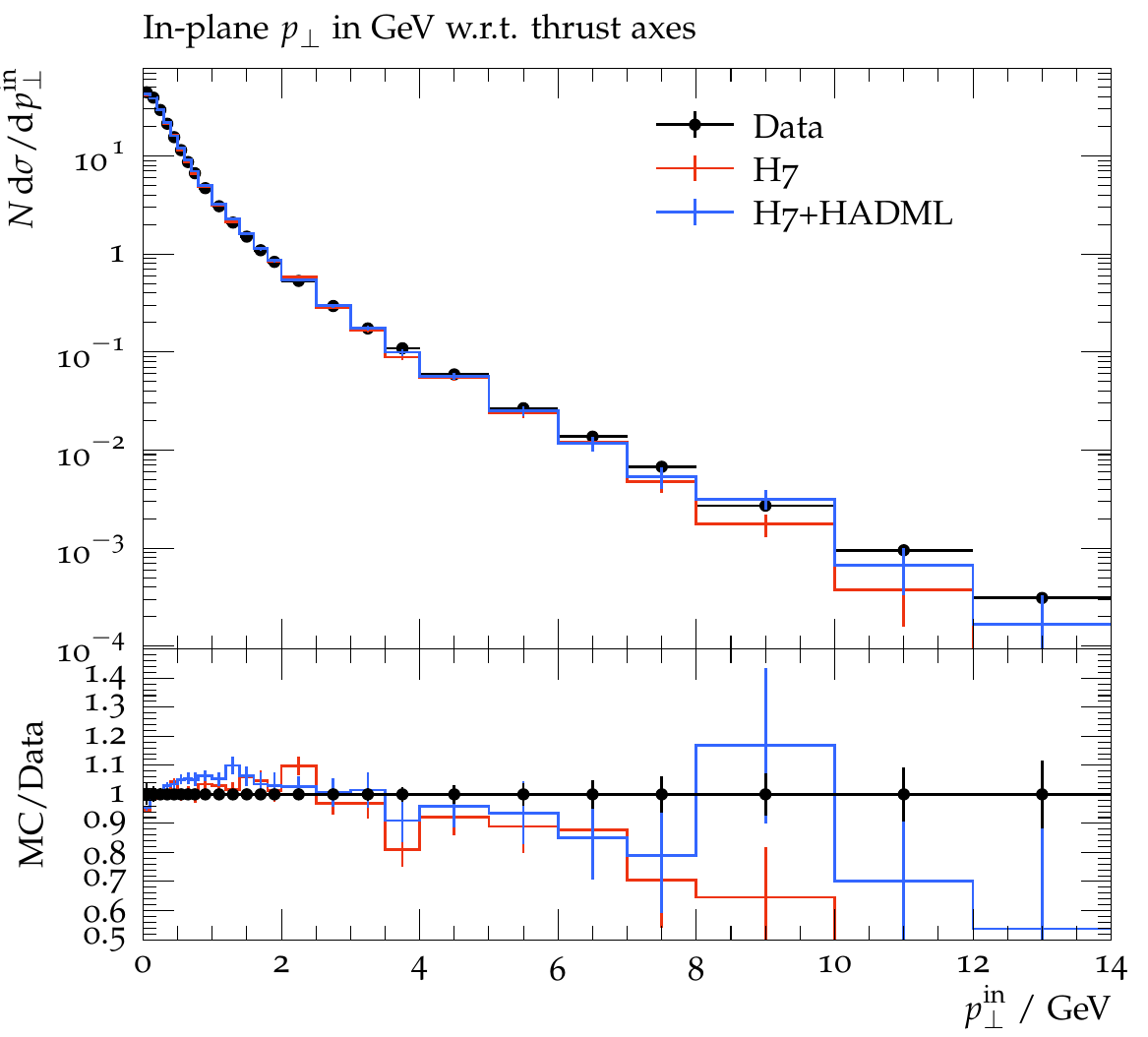}
\includegraphics[width=0.49\textwidth]{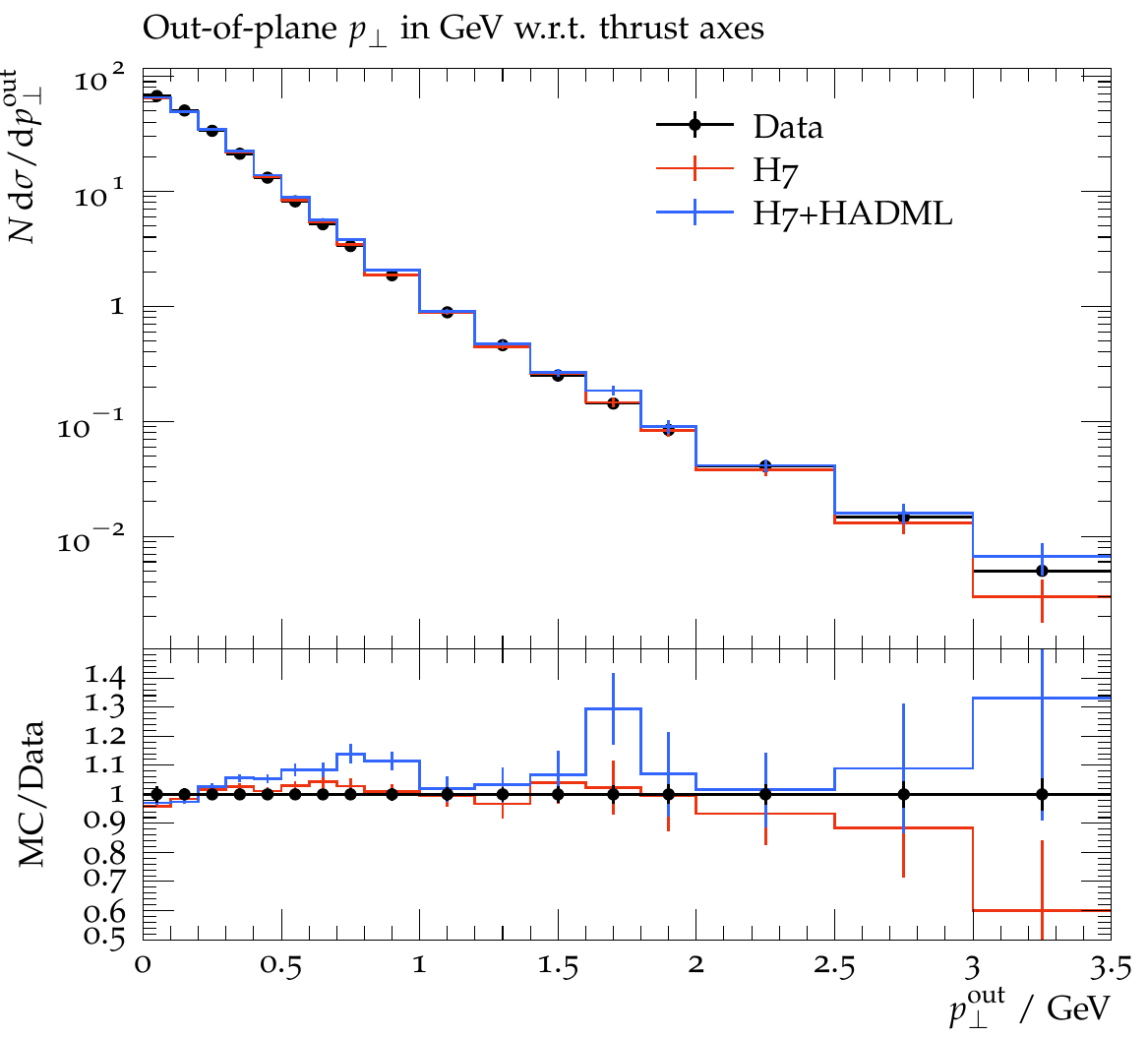}
    \caption{\label{fig:thrust2}  
  Normalized, differential cross-sections of particle transverse momenta along the Thrust major (left) and Thurst minor (right) axes for \HW, \HW\ with our \HADML, and for data from DELPHI at LEP.  Error bars on the predictions represent statistical uncertainties.
    }
\end{figure}

Figure~\ref{fig:thrust} shows histograms of various event shapes.  Thrust~\cite{Farhi:1977sg,Brandt:1964sa} is the quintessential $e^+e^-$ event shape:

\begin{align}
\label{eq:thrust}
    T = \max_{\vec{n}}\left(\frac{\sum |\vec{p}_i\cdot \vec{n}|}{\sum |\vec{p}_i|}\right)\,,
\end{align}
where the sum runs over all final state particle three momenta.  The direction $\vec{n}$ that maximizes the argument of Eq.~\ref{eq:thrust} is called the Thrust axis.  Thrust major is defined similarly to Eq.~\ref{eq:thrust} but with $\vec{n}$ replaced with vectors transverse to the Thrust axis and Thrust minor is the same, but with an optimization only over directions perpendicular to both the Thurst and Thurst major axes.  The Sphericity is computed from the eigenvalues of the quadratic momentum tensor

\begin{align}
\label{eq:spher}
    M^{\alpha\beta}=\sum p_i^\alpha p_i^\beta\,,
\end{align}
where $\alpha,\beta$ are the spatial momentum indices, and the sum runs over the same particles as in Eq.~\ref{eq:thrust}.  Sphericity is defined as $\frac{3}{2}(\lambda_2+\lambda_3)$ for eigenvalues $\lambda_i$ of the $3\times 3$ matrix defined in Eq.~\ref{eq:spher} and $\lambda_3\leq \lambda_2\leq\lambda_1$.  Hadronization shifts event shapes (see e.g., Ref.~\cite{Abbate:2010xh}) and so these observables are sensitive to hadronization modeling.  Figure~\ref{fig:thrust} shows that HADML agrees with \HW\ within 10\% across most of the spectra, which itself agrees with data at a similar level.  Individual particle spectra are shown in Fig.~\ref{fig:thrust2} for the transverse momenta along the Thurst major and minor directions.  The level of agreement is similar to the event shapes where there is sufficient statistical power.

\section{Summary and Outlook}
\label{sec:conclusion}

In this paper, we have established a first step on the path towards a neural network-based hadronization model.  The cluster hadronization model from \HW\ has been emulated with a Generative Adversarial Network.  This model is designed to reproduce the two-body decay of clusters into pions.  The GAN is integrated into the
full \HW\ program by using all other hadronization components from the \HW\ default model.  The kinematic properties of other hadrons are emulated using the pion model and conservation of energy.  We have shown that the \HADML\ is able to reproduce \HW's light cluster decays and when integrated with the full \HW\ simulation, is able to reproduce results from $e^+e^-$ data as well.

The ultimate goal of this research direction is to train the ML model directly on data to improve upon the existing hadronization models.  A number of technical and methodological steps are required to achieve this vision.  First, the deep generative model needs to be extended to directly accommodate multiple hadron species and to model the relative probabilities of the various final states.  In this work, we have modeled different hadron species using conservation of energy, but this means that the fragmentation is assumed universal.  Architectural modifications could allow for perturbations on universality.  Hyperparameter optimization, including the investigation of alternative generative models, is an important component of future work.  Once the deep generative model has the capacity to reproduce all of the physics of the \HW\ cluster model, methodological innovation is required to explore how to tune the model to data.  Traditionally, $e^+e^-$ data are used for tuning.  Optimization with a large set of one-dimensional, binned measurements will need to be explored.  A non-trivial aspect of this optimization is that while the hadronization model would be differentiable, the parton shower input would not be.  Building in a model of uncertainty would also be a central aspect of model tuning.  It may also be possible to tune with unbinned, and higher-dimensional results from $ep$ and $pp$ data~\cite{H1:2021wkz,Arratia:2021otl,Vandegar:2020yvw,Bellagente:2019uyp,1800956,Andreassen:2019cjw}.

While we have focused on hadronization in the context of collider physics, the ideas and concepts described in this paper have broader implications.  First of all, hadronization is used across high energy particle and nuclear physics (see e.g., Ref.~\cite{Sjostrand:2021dal}) and perturbations on the collider model may be required to accurately describe other systems.  Second, there are other physical systems where first-principles input is combined with phenomenological models.  For example, a complete description of observational cosmology requires an $N$-body simulation of the dark matter to be combined with a description of visible matter around dark matter halos (see e.g., Ref.~\cite{2016MNRASFeng,Modi2021A&C,Dai:2020ekz,Bohm:2020ilt,Modi:2021acq}).  While different applications call for domain-specific adaptations, some components and core methodology is common.  Further development in this research area will enable important advances in simulation to improve inference in high energy physics and beyond. \\

\noindent\textbf{Note added}: As this manuscript was being finalized, we became aware of the recent work in Ref.~\cite{Ilten:2022jfm}, which has a similar goal.  That study uses a different Monte Carlo program (\textsf{Pythia} instead of \HW) and uses a different generative model (Variational Autoencoder instead of a GAN). Reference~\cite{Ilten:2022jfm} also focuses on the pion-only case.  
\section*{Acknowledgments}
The work of AS was funded by grant no. 2019/34/E/ST2/00457 of the National Science Centre,
Poland and the Priority Research Area Digiworld under the program Excellence Initiative
– Research University at the Jagiellonian University in Cracow.
BN and XJ were supported by the Department of Energy, Office of Science under contract number DE-AC02-05CH11231. AG is supported by the U.S.\ Department of Energy (DOE), Office of Science under Grant No. DE-SC0009920.




\FloatBarrier
\bibliographystyle{JHEP}
 \bibliography{HEPML,other-refs}

\end{document}